\documentclass{desyproc}

\begin{document}

\title{White dwarfs as physics laboratories: the case of axions}

\author{{\slshape J. Isern$^{1,2}$, 
                  L. Althaus$^{3,4}$,
                  S. Catal\'an$^5$,
                  A. C\'orsico$^{3,4}$,
                  E. Garc\'{\i}a--Berro$^{6,2}$, 
                  M. Salaris$^7$, 
                  S. Torres$^{7,2}$ }\\[1ex]
          $^1$Institut de Ci\`encies de l'Espai ICE(CSIC/IEEC), 
              Campus UAB, 08193 Bellaterra, Spain \\
          $^2$Institut d'Estudis Espacials de Catalunya (IEEC), 
              Ed. Nexus, c/Gran Capit\`a, 08034 Barcelona, Spain\\
          $^3$Facultad de Ciencias Astron\'omicas y Geof\'{\i}sicas, 
              Universidad Nacional de La Plata, Argentina\\
          $^4$CONICET, Argentina \\
          $^5$Center for Astrophysics Research, University of Hertfordshire, 
              College Lane, Hatfield AL10 9AB, UK \\
          $^6$Departament de F\'{\i}sica Aplicada, 
              Universitat Polit\`ecnica de Catalunya,
              c/Esteve Terrades 5, 08860 Castelldefels, Spain\\
          $^7$Astrophysics Research Institute, 
              Liverpool John Moores University, 
              12 Quays House, Birkenhead, CH41 1LD, UK \\ }

\contribID{Isern\_Jordi}

\desyproc{DESY-PROC-2011-04}
\acronym{Patras 2011} 
\doi  

\maketitle

\begin{abstract}
White  dwarfs are  almost  completely degenerate  objects that  cannot
obtain energy from thermonuclear sources, so their evolution is just a
gravothermal cooling process.  Recent improvements in the accuracy and
precision  of  the luminosity  function  and  in  pulsational data  of
variable  white  dwarfs suggest  that  they  are  cooling faster  than
expected from conventional theory.   In this contribution we show that
the  inclusion of an  additional cooling  term due  to axions  able to
interact  with electrons  with a  coupling constant  $g_{\rm  ae} \sim
(2-7) \times 10^{-13}$ allows to fit better the observations.
\end{abstract}

\section{Introduction}

During  the cooling process,  white dwarfs  experience some  phases of
pulsational   instability   powered   by   the   $\kappa$-   and   the
convective-driven   mechanisms   \cite{review}.    Depending  on   the
composition of the atmosphere, variable white dwarfs are known as DAV  
(atmospheres dominated by H) and DOV, DBV  (atmospheres non dominated by H).   
These  objects  are experiencing  $g$-mode  non-radial
pulsations, where  the main restoring force is  gravity.  An important
characteristic of these pulsations is that their period experiences a
secular drift caused by the evolution of their temperature and radius.
For  a semi-qualitative purpose  this drift  can be  well approximated
by~\cite{Winget:1983}: $d\ln  \Pi/dt \simeq -a\, d\ln T/dt  + b\, d\ln
R/dt $,  where $a$ and  $b$ are constants  of the order of  unity that
depend on  the details of the model,  and $R$ and $T$  are the stellar
radius  and  the  temperature  at  the  region  of  period  formation,
respectively.  This equation reflects the fact that, as the star cools
down, the  degeneracy of  electrons increases, the  buoyancy decreases
and, as a consequence, the  spectrum of pulsations gradually shifts to
lower frequencies.   At the same  time, since the star  contracts, the
radius decreases and the frequency  tends to increase. In general, DAV
and DBV  stars are  already so cool  (and degenerate) that  the radial
term is  negligible and the change  of the period of  pulsation can be
directly related  to the change in  the core temperature  of the star.
Therefore, the measurement of such drifts provides an effective method
to test  the theory of white  dwarf cooling. This, in  turn, allows to
obtain a  simple relationship~\cite{Isern:1992,Isern:2008} to estimate
the influence of an additional sink of energy, axions for instance, on
the period drift of variable white dwarfs:

\begin{equation}
({L_{\rm X}}/{L_{\rm model}}) \approx ({\dot{\Pi}_{\rm obs}}/
{\dot{\Pi}_{\rm model}}) -1 
\label{eqf}
\end{equation}

\noindent  where the  suffix ``model''  refers to  those  models built
using standard physics and $L_{\rm X}$ is the extra luminosity.

Another way to test the theory  of white dwarf cooling is based on their
luminosity function.   This function is defined as  the number density
of white dwarfs of a given luminosity per unit magnitude interval:

\begin{equation}
n(l) = \int^{M_{\rm u}}_{M_{\rm l}}\,\Phi(M)\,\Psi(t)
\tau_{\rm cool}(l,M) \;dM
\label{lf}
\end{equation}

\noindent   where  $t$   satisfies   the  condition   $t  =   T-t_{\rm
cool}(l,M)-t_{\rm PS}(M)$ and $l = -\log (L/L_\odot)$, $M$ is the mass
of the parent star (for  convenience all white dwarfs are labeled with
the  mass of  the main  sequence  progenitor), $t_{\rm  cool}$ is  the
cooling time down to luminosity $l$, $\tau_{\rm cool}=dt/dM_{\rm bol}$
is the  characteristic cooling time,  $M_{\rm s}$ and $M_{\rm  i}$ are
the maximum and the minimum masses  of the main sequence stars able to
produce a white dwarf of  luminosity $l$, $t_{\rm PS}$ is the lifetime
of  the progenitor  of the  white dwarf,  and $T$  is the  age  of the
population under  study.  The  remaining quantities, the  initial mass
function, $\Phi(M)$,  and the star formation rate,  $\Psi(t)$, are not
known a priori and depend  on the properties of the stellar population
under study.  In order to  compare theory with observations  and since
the total density of white dwarfs  is not well known yet, the computed
luminosity function is usually normalized to the bin with the smallest
error bar, traditionally the one with $l=3$.  An important property of
Eq.~(\ref{lf}) is that the bright branch of the luminosity function is
only  sensitive to the  average characteristic  cooling time  of white
dwarfs  at   the  corresponding  luminosity  when   this  function  is
normalized.

\begin{equation}
n =  \left\langle\tau_{\rm cool}\right\rangle \int_{M_{\rm l}}^{M_{\rm
u}}\phi(M) \psi(T-t_{\rm cool}-t_{\rm ps})\;dM.
\label{lf1} 
\end{equation}

\noindent The reason~\cite{Isern:2008, Isern:2009} is that the stellar
population is dominated  by low-mass stars and, since  the lifetime of
stars increases very sharply when  the mass decreases, the lower limit
of  the  integral in  Eq.~(\ref{lf1})  is  almost  independent of  the
luminosity,  so  the  value  of   the  integral  is  absorbed  by  the
normalization constant.

\section{The case of G117--B15A and the luminosity function}

The measurement  of the secular drift  of the period  of pulsation has
been performed in the case of G117--B15A ~\cite{Kepler:2005}, a member
of the ZZ Ceti (DAV) stars. The most recent value obtained so far is
\cite{kepler:2011}:

\begin{equation}
(d\Pi/dt)_{\rm obs} = (4.89 \pm 0.53 \pm 1.56) \times 10^{-15}\, {\rm s/s} 
\end{equation}

\noindent  with an estimated  proper motion  correction $  \dot{\Pi} =
-(7.0\pm  0.2)\times 10^{-16}\,  {\rm  s/s}$. Theoretical  predictions
\cite{Corsico:2011}  indicate  that  this  star  should  experience  a
secular drift of only $  \dot{\Pi} = 1.2 \times 10^{-15}\, {\rm s/s}$.
Similar values  ($ \dot{\Pi} = 1.92  \times 10^{-15}\, {\rm  s/s}$ or $
\dot{\Pi} = 2.98 \times 10^{-15}\,  {\rm s/s}$ depending on the adopted
mass   of    the   envelope)   have    been   independently   obtained
\cite{Bischoff:2008}.   These results  suggest that  white  dwarfs are
cooling faster  than expected (it important to confirm this statement by  
measuring this drift in other stars).  There  are three possible  reasons for
this. i) An observational error.  This measurement is difficult and it
has been  obtained by only  one team on  just one star.   Although the
measurement tends to stabilize, it suffered strong fluctuations in the
past  \cite{Isern:2010}.  ii)  A  modelling error.   Models have  been
noticebly improved during  the last ten years and  the two independent
models computed up to  now \cite{Corsico:2011, Bischoff:2008} are in a
qualitative  agreement.    iii)  An  additional  sink   of  energy  is
responsible of the accelerated cooling rate \cite{Isern:1992}.

Under the  conditions of  temperature and density  in the  interior of
G117-B15A,  the   axion  emission   rate  is  dominated   by  electron
bremsstrahlung   (only  the   DFSZ  axion   model)  that   behaves  as
$\dot{\epsilon}_{\rm ax}  \propto g^2_{\rm aee}  T^4_7$ erg/g/s, where
$T_7$ is the temperature in units of $10^7$ K and $g_{\rm aee}$ is the
strength      of     the      axion--electron      Yukawa     coupling
\cite{Nakagawa:1987}.  Thus  it  is  possible  to  include  the  axion
emissivity  in Eq.~(\ref{eqf})  and adjust  $g_{\rm aee}$  to  fit the
observed values. The  value that best fits the  observations is in the
range of $g_{\rm aee} \sim (3-7) \times 10^{-13}$.
 
\begin{figure}[t]
\vspace{6cm}
\includegraphics{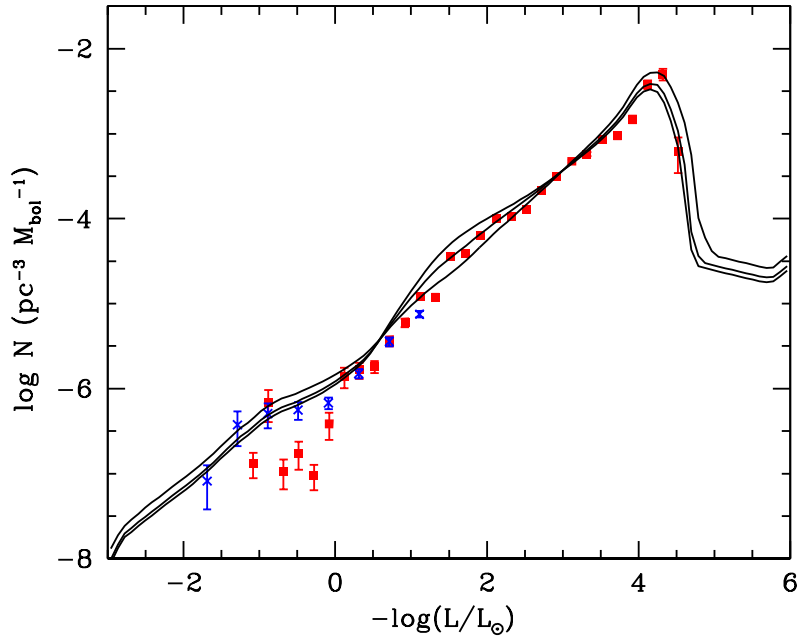}
\caption{White dwarf  luminosity function.  The  solid lines represent
         the models obtained with (up to down) $g_{\rm aee}/10^{-13} =
         0, 2.2, 4.5$ respectively.}
\label{Fig:lum}
\end{figure}

Figure \ref{Fig:lum} displays the  white dwarf luminosity function, DA
and non--DAs  \cite{Isern:1998}. The values  from the DR4 of  the SDSS
\cite{Harris:2006}  (squares) were  complemented at  high luminosities
with those  obtained using the more  recent DR7 \cite{Krzesinski:2009}
(crosses).  It  is important  to  notice  that  models without  axions
\cite{Salaris:2010} predict  an excess of  white dwarfs in  the region
$\log  (L/L_\odot) \sim  -2$  as in  the  case of  the pure-DA  sample
\cite{Isern:2008}. If  axions are included,  the best fit  is obtained
for $g_{\rm aee} \sim  (2-3) \times 10^{-13}$. The luminosity function
at high luminosities  is still poorly known from  both the theoretical
and observational  point of views, but  it is clear  that will provide
strong constrains.

\section{Conclusions}

There are two independent  pieces of evidence, the luminosity function
and  the secular  drift of  DAV white  dwarfs, that  white  dwarfs are
cooling  down  more rapidly  than  expected.  The  introduction of  an
additional sink of energy linked  to the interaction of electrons with
a light boson  (axion, ALP,\ldots) with an strength  $g_{\rm aee} \sim
(2-7)\times  10^{-13}$ solves  the problem  satisfactorily.  Naturally,
the   uncertainties  that   still  remain,   both   observational  and
theoretical, still prevent to claim the existence of such interaction.
A  systematic  analysis aimed  to  discard  any possible  conventional
solution is under way.

\section*{Acknowledgments}

This work has  been supported by the MICINN  grants AYA08-1839/ESP and
AYA2008-04211-C02-01, by the ESF EUROCORES Program EuroGENESIS (MICINN
grant EUI2009-04170),  by SGR grants  of the Generalitat  de Catalunya
and by the EU-FEDER funds.

\begin{footnotesize}

\end{footnotesize}

\end{document}